\documentclass[sigconf, screen, authorversion]{acmart}

\usepackage{multirow}
\usepackage{url}
\usepackage{tabularx}
\usepackage{booktabs}

\copyrightyear{2023}
\acmYear{2023}
\setcopyright{acmlicensed}\acmConference[AIES '23]{AAAI/ACM Conference on AI, Ethics, and Society}{August 8--10, 2023}{Montréal, QC, Canada}
\acmBooktitle{AAAI/ACM Conference on AI, Ethics, and Society (AIES '23), August 8--10, 2023, Montréal, QC, Canada}
\acmPrice{15.00}
\acmDOI{10.1145/3600211.3604679}
\acmISBN{979-8-4007-0231-0/23/08}

\begin{document}

\title{A Systematic Review of Ethical Concerns with Voice Assistants}

\author{William Seymour}
\email{william.1.seymour@kcl.ac.uk}
\orcid{0000-0002-0256-6740}
\affiliation{%
  \institution{King's College London}
  \streetaddress{Bush House, 30 Aldwych}
  \city{London}
  \country{UK}
  \postcode{WC2B 4BG}
}

\author{Xiao Zhan}
\email{xiao.zhan@kcl.ac.uk}
\orcid{0000-0003-1755-0976}
\affiliation{%
  \institution{King's College London}
  \streetaddress{Bush House, 30 Aldwych}
  \city{London}
  \country{UK}
  \postcode{WC2B 4BG}
}

\author{Mark Cot\'{e}}
\email{mark.cote@kcl.ac.uk}
\affiliation{%
  \institution{King's College London}
  \streetaddress{Chesham Building, Strand}
  \city{London}
  \country{UK}
  \postcode{WC2R 2LS}
}

\author{Jose Such}
\email{jose.such@kcl.ac.uk}
\affiliation{%
  \institution{King's College London}
  \streetaddress{Bush House, 30 Aldwych}
  \city{London}
  \country{UK}
  \postcode{WC2B 4BG}
}


\begin{abstract}
Since Siri's release in 2011 there have been a growing number of AI-driven domestic voice assistants that are increasingly being integrated into devices such as smartphones and TVs. But as their presence has expanded, a range of ethical concerns have been identified around the use of voice assistants, such as the privacy implications of having devices that are always listening and the ways that these devices are integrated into the existing social order of the home. This has created a burgeoning area of research across a range of fields including computer science, social science, and psychology. This paper takes stock of the foundations and frontiers of this work through a systematic literature review of 117 papers on ethical concerns with voice assistants. In addition to analysis of nine specific areas of concern, the review measures the distribution of methods and participant demographics across the literature. We show how some concerns, such as privacy, are operationalized to a much greater extent than others like accessibility, and how study participants are overwhelmingly drawn from a small handful of Western nations. In so doing we hope to provide an outline of the rich tapestry of work around these concerns and highlight areas where current research efforts are lacking.
\end{abstract}

\maketitle

\section{Introduction}
The last decade has seen the widespread introduction of voice assistants (VAs) into domestic life in many parts of the world. Offering novelty and convenience driven by advancements in AI technologies such as machine learning and natural language processing, VAs have transformed the home computing landscape. Positioned by vendors at the centre of the smart home as a hub for other apps and gadgets, research shows that VAs are most commonly used to play music, search for information, and control other IoT devices \cite{10.1145/3311956}. In this way, their usage extends that of the smartphone where app stores allow for the use of a wide variety of third-party software, and many smartphone apps are also available as skills/actions. But VAs do not simply offer access to traditional means of computing via a new interaction modality, their design and interfaces also represent a number of novel experiences and changes in people's underlying relationship with the technology that they use.

The continued integration of speech into the smart \textit{home}---a space often idealised as private, safe, and intimate---disrupts existing social norms, and the frequent gendering of voice assistants as female has prompted harsh criticism of the way that VAs implicitly perpetuate stereotypes around gendered work~\cite{west2019blush}. After police took steps to use Alexa recordings in a murder trial, legal scholars began to examine more seriously the (lack of) protections for data that is collected in the home but stored in the cloud~\cite{pfeifle2018alexa}. In some cases this represents the latest in ongoing debates around parenting and privacy as VAs challenge and reframe existing norms by altering what is and is not possible. In other areas VAs have resurrected much older ethical concerns, revealing new dimensions of long-standing concepts like anthropomorphism. While the affective potential of computers that use natural language has been known for decades~\cite{picard2000affective, weizenbaum1966eliza}, voice assistants take this previously unattainable technical capability of \textit{conversation}---that activates the same areas of the brain as speech between people~\cite{nass2005wired}---and scales it to billions of speakers, TVs, headphones, smartphones, and other devices around the globe.

The breadth of these ethical concerns means that research has emerged from a diverse range of disciplines, including computer science, social science, and psychology, each with different practices and conventions. The sensitive nature of the home environment and relationships drives us now to pause and take stock of the literature on ethical concerns with voice assistants. To this end we conducted a systematic literature review with the aim of capturing the concerns that have been identified and how they are studied. The results are valuable both in understanding current areas of enquiry as well as in identifying opportunities for future work. Beyond this we were also interested in the diversity and inclusion of participants. When conducting research that tells us about people and their social interactions we must acknowledge that people of different cultures, ages, genders, abilities, etc. experience voice assistants differently and have different concerns about them. As a provocation intended to foster a more inclusive---and accurate---body of knowledge, we use our review to highlight the overwhelming bias towards ``WEIRD'' (Western, Educated, Industrialized, Rich, and Democratic) countries in the venues searched. Beyond geography we also examine other dimensions of diversity in research that focuses on specific groups, such as those around gender, that reflect and inform social and cultural norms.

More specifically we answer the following research questions:
\begin{enumerate}
\item[RQ1] What is the current state of research on ethical concerns in voice assistants?
\item[RQ2] How are participants, methods, and approaches represented across this research?
\end{enumerate}

\noindent And in so doing make the following contributions:
\begin{itemize}
    \item Map out the current knowledge on ethical concerns around privacy, social interaction, accessibility, social order, performance of gender, accountability, conflicts of interest, misinformation, and transparency
    \item Show that research on voice assistants overwhelmingly studies WEIRD demographics
    \item Highlight key directions for future work in this area, including challenging legacy assumptions about how VAs are designed and deepening explorations of how VAs interact with gender in society
\end{itemize}

\section{Methodology}
In order to assess prior work on ethical concerns around the design and use of voice assistants in the home we conducted a systematic literature review, i.e., one with ``a clearly formulated question that uses systematic and explicit methods to identify, select, and critically appraise relevant research''~\cite{moher2009preferred}. We followed established guidance on systematic reviews which lays out the main steps around searching for and analysing prior work. This involves identifying: (1) eligibility criteria for included papers; (2) databases to be searched; (3) parameters for the search; (4) the process for selecting studies from the set of returned papers; and (5) how data will be extracted from those papers~\cite{moher2009preferred}.\footnote{As the PRISMA guidelines are intended for clinical reviews, we have omitted steps that would be inappropriate for the present research (e.g., summary measures used across the reported papers).}

\subsection{Eligibility Criteria and Databases Searched}
We considered journal articles, conference papers, extended abstracts, and short papers about voice assistants since 2012 to coincide with the initial commercial availability of voice assistants. As our focus was on voice assistant research, we searched the ACM Digital Library, IEEE Explore, Web of Science, and DBLP.

\subsection{Search Parameters}
Conducting the search presented a `cold start' problem: reviewing every paper on voice assistants to distil out concerns was infeasible, but at the same time there was no existing literature mapping out ethical concerns with VAs from which we could draw keywords to narrow down the search. Following common practice in studies of this type (e.g.~\cite{10.1145/3469886}) we used adjacent prior work and domain knowledge to build a list of keywords. The positioning of VAs as `smart devices' means that study of them often falls under the umbrella of smart home research, so we selected concerns from four papers summarising ethical concerns in smart homes that were applicable to voice assistants~\cite{edwards2001home, wilson2015smart, seymour2020strangers, nilsson2019breaching}. We then drew on domain knowledge to supplement this with ethical challenges specific to VAs using work on individual concerns, highlighting additional issues that arise outside of the main discussion on challenges and barriers to smart home adoption (e.g. around social interaction and the performance of gender). During this process we adopted a broad view of ethics and related concerns; following prior work we define ethics as ``what a design object ought to be based on ethical and moral codes'', in contrast with its purpose/function (reason), and visual values/presentation (aesthetics)~\cite{10.1145/3173574.3173578}. The final set of resulting keywords is given in Table~\ref{tab:criteria}. During the review we adjusted the categorisation of concerns to best describe the literature returned by the survey (more information on this is given in Section~\ref{sec:discoveries}), arriving at the following concerns from papers on smart homes:

\begin{itemize}
    \item Privacy~\cite{edwards2001home, wilson2015smart, seymour2020strangers, nilsson2019breaching}
    \item Agency and Autonomy~\cite{seymour2020strangers, wilson2015smart, nilsson2019breaching,such2017privacy}
    \item Social Order and Accountability~\cite{edwards2001home, seymour2020strangers}
    \item Transparency~\cite{nilsson2019breaching, seymour2020strangers,such2017privacy}
    \item Conflicts of Interest \& Datafication~\cite{nilsson2019breaching, seymour2020strangers}
\end{itemize}

Supplemented by four concerns unique to voice assistants:

\subsubsection{Social Interaction}
The use of speech and conversation by VAs has raised concerns about how they might change how people interact both with them and each other. Work in this space has shown how people automatically apply social rules and draw upon gender stereotypes in interactions with computers~\cite{nass1994computers}, and use anthropomorphism as a heuristic to help develop mental models of computers and robots~\cite{zawieska2012understanding}.

\subsubsection{Performance of Gender}
Popular voice assistants are explicitly gendered: Alexa reports to be ``female in character'', Google Assistant was described by an engineer as ``a young woman from Colorado'', and Siri is a Scandinavian female name~\cite{west2019blush}. What is now an industry norm has been criticised for reinforcing existing societal biases around the role of women in the workforce, portraying them as ``obliging, docile and eager-to-please helpers''~\cite{west2019blush}.

\subsubsection{Accessibility}
Voice assistants present unique challenges and opportunities for accessibility in the smart home. On the one hand, by using voice as their primary or only mode of interaction they align well with the needs of communities such as the blind and partially sighted~\cite{10.1145/3368426}, but as a direct consequence, they disproportionately fail people with speech, language, or hearing difficulties.

\subsubsection{Misinformation}
In an extension of studies around the quality and potential bias of information provided by internet search engines~\cite{haim2018burst}, researchers have begun to examine the information provided by voice assistants~\cite{dambanemuya2021auditing}. This is particularly important for VAs because succinctly conveying the source and accuracy of information provided via speech is a significant challenge.

\subsection{Study Selection}
To be considered in scope, papers had to feature significant results that directly addressed one or more ethical concerns around voice assistants used for domestic tasks in the home or via a smartphone (\textbf{e.g. a study on smart homes would only be in-scope if it had findings specific to voice assistants}). Papers where the results only detailed \textit{solutions} to ethical concerns were out of scope (i.e. where the understanding of concerns came solely from background literature), as were papers that developed or applied voice assistants to tasks or contexts outside of normal domestic use (e.g. medical treatment). Energy usage was not considered an ethical concern in the context of the review as it relates to the smart \textit{home} in general rather than voice assistants as a class of devices. While many papers explicitly listed the keyword topics as concerns, we did not exclude papers that made no or implicit references to the key words as concerns (e.g. participants describing an emotional connection to a voice assistant with no associated normative judgement).

\begin{table}
    \centering
    \begin{tabularx}{\columnwidth}{X|X}
    \toprule
    \multicolumn{2}{c}{Papers containing at least one of the following devices:} \\ \hline
    Alexa & Siri \\
    Google Assistant & Voice assistant \\
    Virtual assistant & Intelligent personal assistant \\
    Smart Home & - \\ \hline
    \multicolumn{2}{c}{And at least one of the following key words:} \\ \hline
    Privacy & Anthropomorphism \\
    Autonomy & Personification \\
    Children & Gender \\
    Conflict of interest & Agency \\
    Social Order & Accessibility \\
    Ethics & Accountability \\
    Transparency & - \\ \hline
    \multicolumn{2}{c}{As an article, conference paper, extended abstract} \\
    \multicolumn{2}{c}{ or short paper published since 2012} \\
    \bottomrule
    \end{tabularx}
    \caption{Search criteria}
    \label{tab:criteria}
    \vspace{-5mm}
\end{table}

\subsection{Data Extraction}
The full text of the selected papers were coded for methodology, type of contribution, and the ethical concern(s) targeted. Two researchers initially considered a small subset of the papers, meeting to compare and refine the process before applying this to the rest of the data. To the end of answering RQ2 on the diversity of those who are represented in human-AI interaction (HAI) and human-computer interaction (HCI) research, we also coded the geographic location of participants (i.e. country of residence). Where participants from multiple countries were sampled, papers were coded to the majority demographic. In a small number of cases where papers listed the platform used for recruitment without specifics on participants' country of residence, those papers were coded with the majority for that platform (in all cases this was the US).

\subsection{Results of the Search Process}
\begin{figure}
    \centering
    \includegraphics[width=0.9\columnwidth]{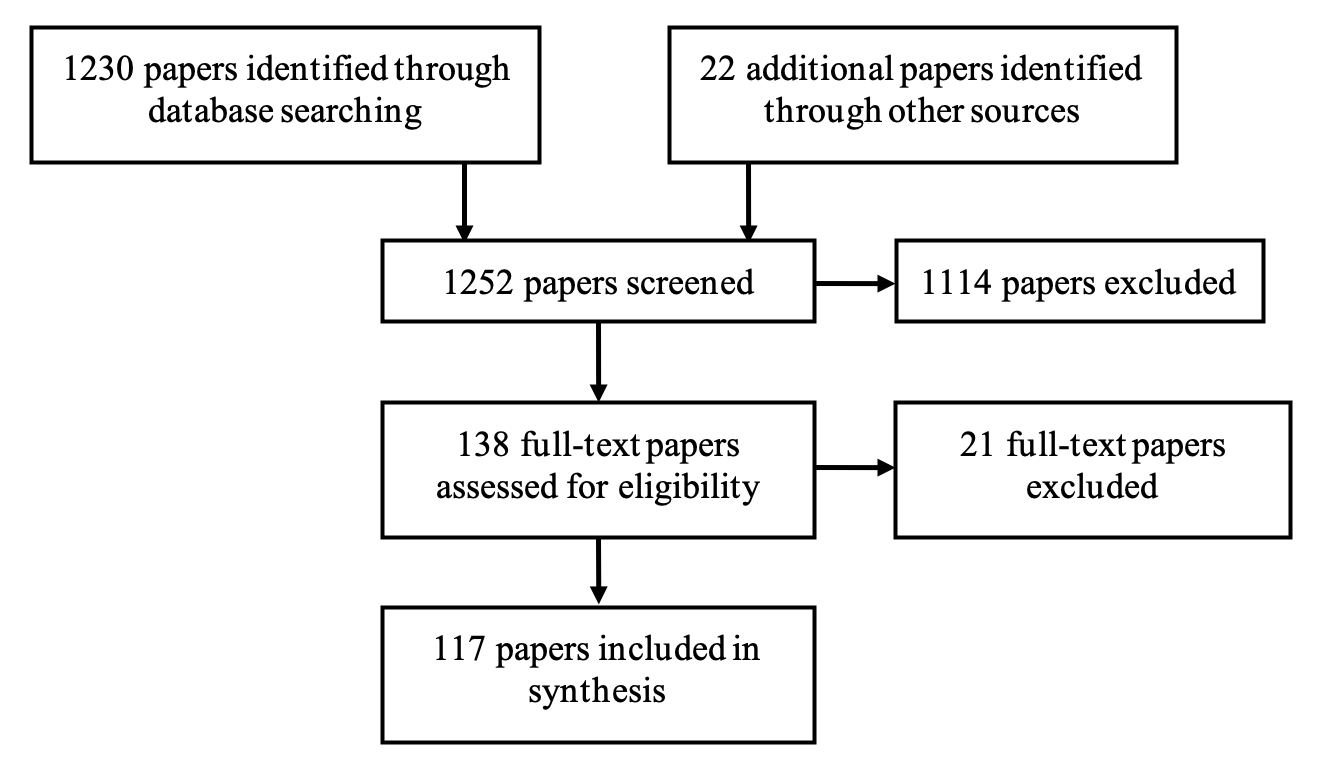}
    \caption{Flow diagram of the search process.}
    \label{fig:screening}
\end{figure}

The initial keyword search returned 1230 unique papers (ACM: 384, IEEE: 389, Web of Science: 266, DBLP: 191). While the systematic search was generally very effective, a small number of relevant papers (22) known to the research team fell outside the range of the search, and were also added. This occurred mainly because the papers were not indexed by the chosen platforms, and occasionally because they used non-trivial variations of the search keywords (e.g. social cohesion/group dynamics vs social order in \cite{lee2020hey}). The first sift identified 138 papers as potentially within scope, which were then coded for methodology, type of contribution, participant country of residence, and the ethical concern(s) targeted. The papers were then grouped by primary concern, and the analysis below is the result of repeated iterations by the wider research team. Four papers were excluded at this stage due to being unobtainable online and 17 for being outside the scope of the review, leaving 117 papers for full analysis. A flow diagram of the search process is given in Figure~\ref{fig:screening}, a record of the included papers and categorisations is available online at \url{https://osf.io/p4h2r}, and numerical overviews of the review categories are provided in the Appendix.

\section{Research Trends}\label{sec:statistics}
\subsection{What Concerns are Studied, and How?}\label{sec:whatconcerns}
\begin{figure}
    \centering
    \includegraphics[width=0.9\columnwidth]{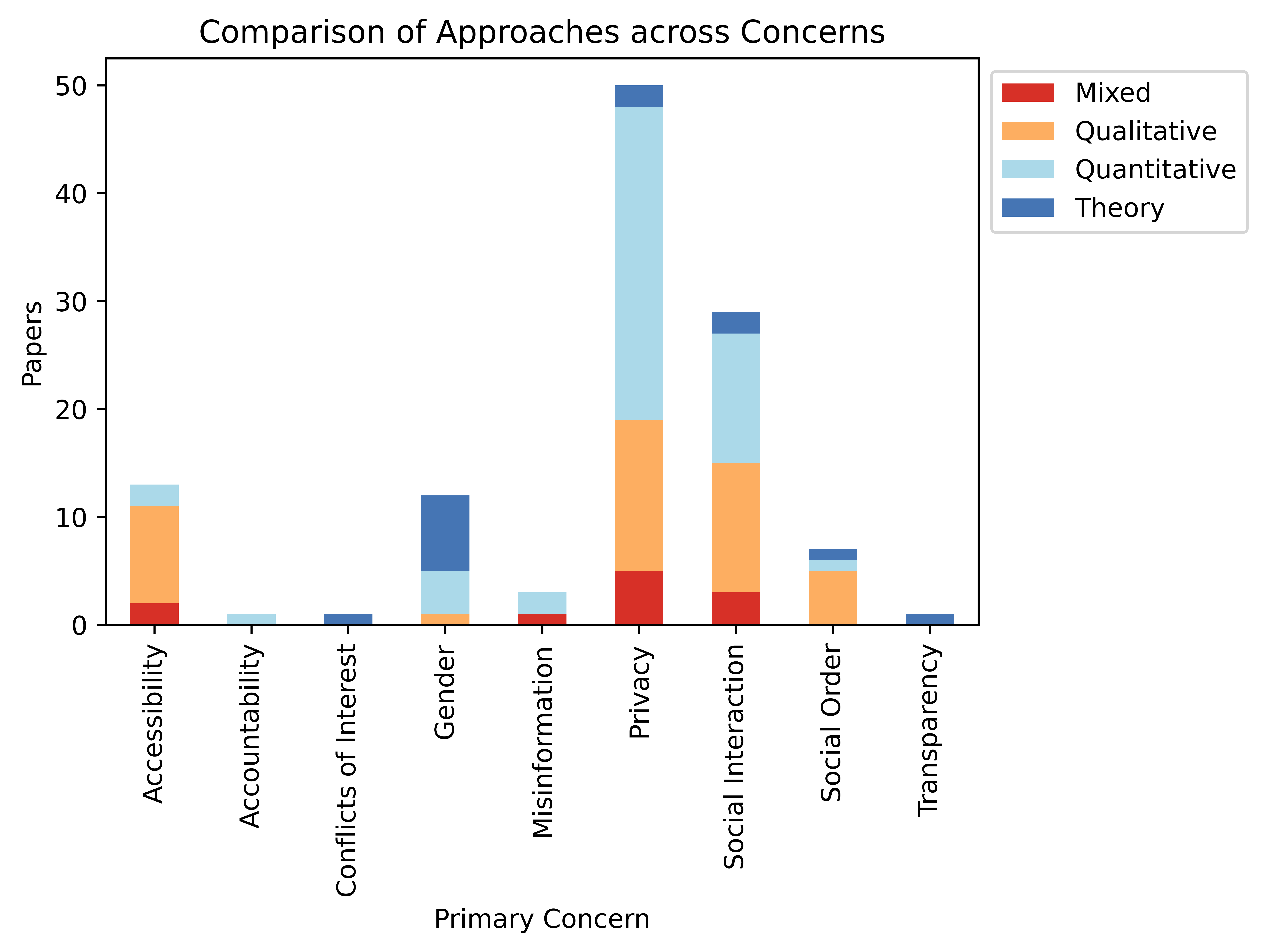}
    \caption{Approaches used by primary concern.}
    \label{fig:approachbyconcern}
\end{figure}

\begin{figure}
    \centering
    \includegraphics[width=0.9\columnwidth]{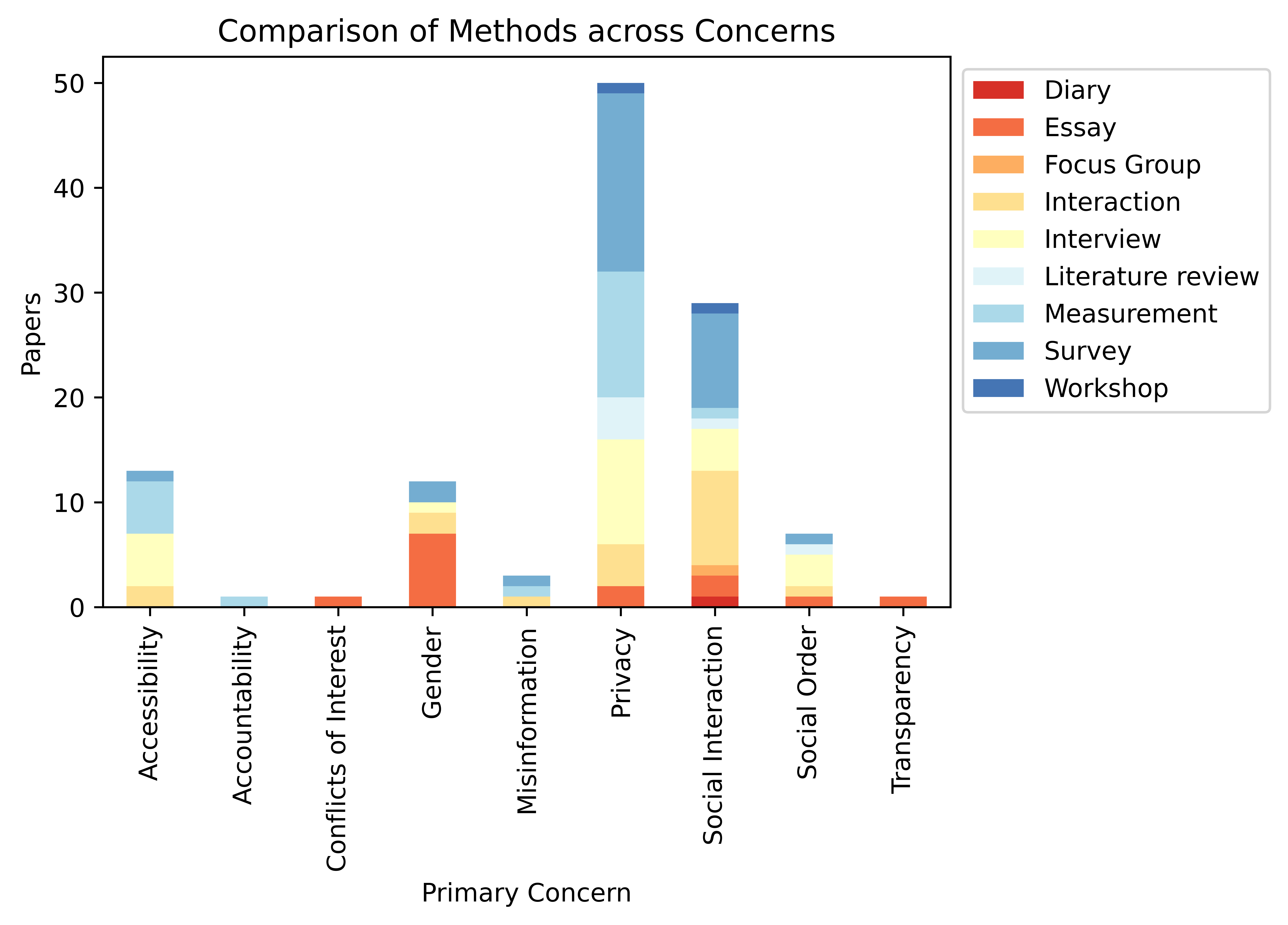}
    \caption{Methods used by primary concern. Note that papers with similar methods may have their research approaches (e.g. qualitative, quantitative, etc.) coded differently depending on how the analysis was conducted.}
    \label{fig:methodbyconcern}
\end{figure}

This section describes the distribution of approaches and methods across the reviewed papers in order to answer RQ2. Unsurprisingly, privacy was the most prevalent concern investigated, followed by social interaction. The high-level research approaches adopted show quantitative methods as most common followed by qualitative, theoretical, and finally mixed approaches. Overall there was a greater diversity in methods amongst qualitative approaches, although surveys and interviews together represented approximately 39\% of research methods. Figures~\ref{fig:approachbyconcern} and~\ref{fig:methodbyconcern} show the relative proportions of concerns, approaches, and methods in the review sample. A full breakdown is provided in the appendix.


From the charts it is clear that several ethical concerns appear under-researched given current public debate around voice assistants and digital technology more generally. Misinformation and performance of gender stand out as particular examples of this, although the latter \textit{was} the centre of burgeoning discussion---6 out of the 14 theoretical papers in the review discussed the performance of gender by VAs, and these are likely to form the foundation of future empirical studies. Related to this it appears that certain concerns are perceived to be more readily operationalised than others. While 67\% of empirical privacy research utilised a quantitative approach over a qualitative one, for empirical research on social interaction both were equally likely, and empirical research on accessibility was \textit{much }more likely to be qualitative (82\%). 

Our understanding of all of the concerns described in the review, even the more extensively researched such as privacy, is still evolving, and prematurely quantifying them can lead to research losing sight of the users (people) behind the voice assistants we study. Related to debates within the fairness, accountability, and transparency community on how appropriate it is to quantify various aspects of the human experience that then feed into vast assemblages of data processing~\cite{10.1145/3442188.3445901, 10.1145/3433949}, researchers should be cautious when attempting to quantify the lived experiences that relate to the concerns described in the review. The analysis shows that researchers are much more comfortable doing this for some concerns than for others, and this leads to the literature giving a distorted view of how best to investigate different concerns. When ethical concerns are quantified it invites comparison between the resulting metrics, whether this is intentional or not, which is problematic when definitions vary between data sets; efforts to create privacy labels demonstrate this tension well by often equating the nuanced personal concerns of individuals with abstract scores representing technical device behaviours~\cite{10.1145/3290605.3300764}. This occurs despite a more complex understanding of privacy outside of the voice assistant literature: \citeauthor{nissenbaum2020privacy}'s model of privacy as contextual integrity, for example, heavily involves norms and relationships~\cite{nissenbaum2020privacy}. The `Seven Veils' model of privacy similarly includes rights and phenomenological aspects that encapsulate different meanings of privacy and clearly benefit from a deep understanding of the human experiences involved~\cite{ohara2016seven} that we argue should not be primarily investigated via a quantitative lens.

On the other hand, there clearly are some situations where quantitative approaches are required, such as in studies of the ecosystems that underpin voice assistants through skill/action marketplaces. Here automated approaches are the only or most efficient means to scale analysis for work on many thousands of skills or reviews. Similarly, quantification may be more appropriate for research questions that follow an established model or framework and use terms related to these concerns in a well-defined way.

Despite the 10 year search window the oldest paper in the literature review was from 2015, with 87\% of papers published in the past three years. The most numerous year was 2020, comprising 37\% of the sample. Comparing research methods and approaches from 2019 to 2021 reveals an increase in the proportion of publications utilising quantitative methods from around 41--42\% in 2019 and 2020 to 63\% in 2021,\footnote{The review snapshots were taken between September and November of 2021, meaning that papers indexed after that time are not included in the results.} likely as research moved online during the coronavirus pandemic.

\subsection{Whose Voices are Heard in VA Research?}\label{sec:whoconcerns}
An ongoing problem in HAI is the lack of diversity amongst the people who are included in research. With this in mind, the community has so far sought to create a welcoming and accessible space for researchers and practitioners. However, the results of the analysis indicate that diversity amongst the \textit{participants} in HAI research remains a major problem. When answering RQ2, 94\% of the 85 papers were conducted solely or mostly on participants from North America and Europe (see Table~\ref{tab:geomap}). This mirrors a long-running trend in HCI research, with 73\% of CHI studies from 2016--2020 recruiting participants from nations with less than an eighth of the world's population~\cite{10.1145/3411764.3445488}. Our review of papers across a range of HAI and HCI venues shows that this effect is even more pronounced in voice assistant research. During this analysis we were forced to reflect on our own prior research, which exhibits the very same `WEIRD' biases. This is something that we aim to address in future research projects, and hope that others will join us in creating a more inclusive global account of how voice assistants are perceived and used.

Another aspect of participation included in the analysis was the specific populations targeted by research studies. Of the 22 papers that explored the experiences of a particular group, 13 were focused on age, 7 on (dis)ability, and 1 each on trans/non binary and Portuguese speakers (see Table~\ref{tab:subpops}). Together these represented just over one quarter of the studies that recruited human participants. Within the papers concerned with age there was a diversity of concerns across anthropomorphism (5), accessibility (4), social order (2), and accountability (1). In contrast, all of the papers that recruited based on ability were focused on accessibility concerns. Studies looking at specific groups mainly took qualitative approaches (70\%), suggesting a focus on lived experiences rather than the gathering of a high level overview. This may present an opportunity for future work (e.g. the state of accessibility across an entire platform).

\begin{table}
    \centering
    \begin{tabular}{lc|lc}
    \toprule
    Country & Papers & Country & Papers \\
    \midrule
    USA & 47 & Netherlands & 2 \\
    UK & 11 & Ireland & 1 \\
    Germany & 10 & South Korea & 1 \\
    Italy & 4 & Spain & 1 \\
    India & 2 & Sweden & 1 \\
    Brazil & 2 & Switzerland & 1 \\
    Canada & 2 && \\
    \bottomrule
    \end{tabular}
    \caption{Participant country of residence for papers with human participants.}
    \label{tab:geomap}
    \vspace{-5mm}
\end{table}

\section{Analysis of Ethical Concerns}
\subsection{Privacy} \label{sec: privacy}
As with many other areas of research, voice assistant privacy is a concept that encompasses a variety of perspectives. At a high level, most papers aligned along two approaches: eliciting user concerns over privacy when using voice assistants, and measuring an aspect of voice assistant behaviour in order to make claims about the (lack of) privacy offered by those devices.

A major theme amongst exploratory studies about user concerns was data collection and processing by vendors and third parties, including from the use of voice assistants by inhabitants of the home beyond the primary user. Various distinctions were given to illustrate the boundaries of acceptable data use, including between first/third parties~\cite{10.1145/3274371}, normal/sensitive subject matters~\cite{10.1145/3311956}, and whether data was subsequently analysed by humans or machines~\cite{10.1145/3449119}. Where other entities were mentioned, these were mostly the government~\cite{pridmore2020personal}, or just a vague `other' with malicious intent~\cite{10.1145/3274371}. This was sometimes associated with data breaches of first~\cite{10.1145/3274371} or third parties~\cite{10.1145/3369807}.

Another dominant theme was that of uncertainty when users characterised privacy risks. While the capability of voice assistants to constantly listen to their surroundings was well understood, participants expressed a lack of understanding about when voice recordings were stored by their devices and how this data was subsequently used by vendors. While several papers called for more transparency around data collection and some users were open to having their concerns addressed by vendors~\cite{10.1145/3428361.3428469}, it is not clear if this would be convincing given widespread scepticism over the honesty of major players given their vested interests~\cite{10.1145/3274371}. This uncertainty was complicated by phantom activations~\cite{10.1145/3449119} and confusion over whether the mute button affected the device's microphone or speaker~\cite{10.1145/3274371}.

The root of these concerns was often vague or incorrect mental models of devices~\cite{abdi2019more,10.1145/3313831.3376529}, as demonstrated by poor knowledge of available privacy controls (e.g. the ability to replay and delete stored voice recordings)~\cite{10.1145/3311956}. Where voice profiles were set up they often had high error rates~\cite{10.1145/3313831.3376529}, reducing the perceived reliability of the underlying technology. This often resulted in disengagement through what were sometimes called `informal' coping mechanisms such as unplugging the device~\cite{abdi2019more} or making sensitive requests through other means~\cite{10.1145/3311956}. It is also reflected in the reasons that participants gave for not taking action and continuing to use their devices, such as having nothing to hide~\cite{10.1145/3449119}, being unable to escape the influence of large technology companies~\cite{10.1145/3274371}, disbelief that companies could store the required volume of data~\cite{10.1145/3274371}, or relying on protection from laws and regulations~\cite{10.1145/3449119}. Though participants knew that devices could infringe on their privacy, this was often framed as a necessary tradeoff against functionality or convenience~\cite{10.1145/3274371, 10.1145/3313831.3376529, 10.1145/3369807, 10.1145/3170427.3188448}.

It is clear that user understanding of voice assistants continues to be a key issue, with privacy being a more visible situation where inaccurate mental models come to the fore. This concern has two related components: on-device affordances around recording do not give users sufficient certainty about the device's operation, and users simultaneously lack trust in manufacturers; `soft' mechanisms used to signal device state (e.g. for a muted microphone) are unconvincing in devices designed to constantly listen, and the poor track records of vendors around privacy and ethics creates a situation where these mechanisms may \textit{never} be sufficient because they are unverifiable and rely on trust. It is therefore likely that effective solutions in this area will rely on local processing, the use of open source software, or external devices that limit data collection capabilities.

A more tangible aspect of VA use concerned the tensions and risks posed by other users of shared devices. Cohabitants often overheard interactions with voice assistants~\cite{10.1145/3313831.3376529} and these concerns frequently blurred into more specific concerns around the ability to purchase items via voice command, which was often disabled as a result~\cite{abdi2019more}. More generally, VAs were seen to artificially create or remove asymmetries around access to data that were already established between cohabitants (see also Section~\ref{sec:social_order} on social order concerns). This was typically noticed when requests for content were actioned through someone else's account~\cite{10.1145/3274371} or upon reflection that guests/children would be able to make identical requests~\cite{10.1145/3313831.3376529}. There was also discussion of how both vendors and the law placed the burden on users to manage the privacy of shared devices, and how it was difficult for `bystanders' to effectively opt-out of being recorded~\cite{10.1145/3449119}. Differing mental models of voice assistants by participants led to frequent worries that they or someone else might overstep privacy boundaries, but in many cases the examples given would not be possible on current devices or could be avoided by using personalised voice profiles. Another attack vector that participants expressed concern about was neighbours or other people outside the home's inhabitants giving commands to the voice assistant from outside or while in the home as a guest~\cite{10.1145/3428361.3428469, abdi2019more, 10.1145/3313831.3376529}, and one measurement paper verified the efficacy of this attack~\cite{mccarthy2020shouting}. Motives were typically vague or absent when discussing these instances of `shouting through letterboxes', especially given people's propensity to disable voice-based purchasing. Notably, no papers directly explored the possibility of using speaker differentiation to \textit{prevent} use of a VA. In general, more work needs to be done around the use of devices by multiple users. Current architectures often reflect the requirements of data protection regulations that intentionally operate to the boundary of the home and no further, and thus poorly facilitate e.g. guests and cohabitants.

Quantitative work mostly identified devices, skills, or users from encrypted network traffic, using machine learning techniques to classify packets with generally good results~\cite{10.1145/3427228.3427277, naraparaju2020fingerprinting, kalin2019amaze, 10.1145/3320269.3384732, kennedy2019can}. User facing studies included perceptions of voice recording~\cite{10.1145/3339252.3340330} and data sharing~\cite{10.1145/3411764.3445122}, as well as the effectiveness of privacy policies~\cite{10.1145/3427228.3427250}. Results supported qualitative findings that the privacy aspects of these devices are generally poorly understood by users. Finally, there was a cluster of work on attack surfaces~\cite{10.1145/3412383}, including through forensic examination of local files and data accessed via APIs~\cite{krueger2020using, germanos2020privacy}. This was linked to the discussion around the influence of the GDPR and existing or desired legal protections for voice assistant data. In general discussions around values and regulation were exclusively focused on the West, with only one paper acknowledging this bias in current research~\cite{pridmore2020personal}).

\subsection{Social Interaction}
Work on social interaction covered a wide range of concepts and phenomena. Originally styled as anthropomorphism when planning the search terms, these works also covered ontological explorations of the separability of humans and machines and qualitative accounts of relationships and emotional connections with VAs. The challenge with mapping these facets of social interactions was the extent to which they overlaped; affective experiences are connected to but distinct from emotional connection, and personification is related to anthropomorphism as similes are to metaphors. The theoretical foundations of these papers show a similar diversity, commonly building on the Computers are Social Actors (CASA) paradigm (e.g.~\cite{lee2021social, 10.1145/3027063.3053246, 10.1145/3290605.3300488}), but other theories and models from HCI~\cite{10.1145/3404983.3405513}, psychology~\cite{whang2021like}, and communications theory~\cite{10.1145/3479515} feature widely in the included papers.

A strand of research around the causes and effects of anthropomorphism designed, adapted, and reused question inventories measuring a wide range of social and behavioural constructs. This was typically used to develop a model that explained variations in one factor using the others. Of concern here was also how these effects might be (mis)used in different contexts~\cite{10.1145/3290607.3310422}, as social interactions with voice assistants could be seen to work alongside or orthogonally to more logical means of decision making.

Work on the fundamental nature of voice assistants attempted to understand how they fit into existing understandings of technology and social interaction. While there were theoretical contributions in this area there were also many empirical investigations into people's responses to increasingly social devices (e.g.~\cite{10.1145/3313831.3376665, 10.1145/3404983.3405513, moussawi2020perceptions}) and what the nature of the relationships they form with them might be~\cite{10.1145/3359316, ghosh2020understanding, 10.1145/3479515}. A key theme here was one of categories; ontologically speaking, do people think of voice assistants as people or machines? Studies addressing this often focused on children who were still learning to understand the world, observing how they responded to stimuli from both people and devices including VA-originated social interactions such as displays of friendship or praise~\cite{10.1145/3383652.3423906, festerling2020alexa, aeschlimann2020communicative}. \citeauthor{festerling2020alexa} ask whether VAs might cause the emergence of a new intermediate category (the New Ontological Category Hypothesis~\cite{10.1145/1957656.1957710}), with ultimately inconclusive evidence; young participants engaged with compliments from VAs as if they were real (i.e. rather than pretend-play), but often attributed advanced features of VAs to `human interference' in an attempt order to keep categories pure~\cite{festerling2020alexa}. Other studies show that children cooperate differently with VAs and humans, such as sharing updates with human collaborators but recognising that voice assistants ``do not care much about progress talk''~\cite{aeschlimann2020communicative}. This also holds true in adults, with relationship development metrics coalescing into a single group that might suggest ongoing purification work (redefining category boundaries to place voice assistants firmly in one or the other)~\cite{10.1145/3479515, 10.1145/2470654.2466455}. \citeauthor{10.1145/3359316} further found that participants fluidly moved \textit{between} categories depending on the behaviour of their voice assistants, treating them as a person during social interaction but as an object at other times~\cite{10.1145/3359316}.

Another recurring phenomenon in empirical work was the use of gendered pronouns as a means of demonstrating or measuring anthropomorphism. People fluidly shift between the use of gendered and impersonal pronouns~\cite{10.1145/3359316} similarly to the shifts between categories described above, with impersonal pronouns making up the majority of references~\cite{10.1145/3027063.3053246}. These types of responses were often attributed to over-learned, reflexive social routines rather than meaningful displays of social intimacy~\cite{10.1145/3176349.3176868}, supported by observations that people often continue these personifying behaviours after learning more about VAs and how they work~\cite{10.1145/3459990.3460730}. There was also a general understanding that social interactions with voice assistants are in some sense inherently satisfying~\cite{moussawi2020perceptions, 10.1145/3027063.3053246}, with explanations often focusing on the similarity between these interactions and those found in interpersonal social situations (i.e. the familiarity of conversation)~\cite{10.1145/3308532.3329466}. There is a balance to be struck between the way that the underlying satisfaction from engaging socially with a device can mask or ameliorate its other drawbacks~\cite{benlian2020mitigating} and the way that appearing `too human' can instil unattainable expectations of a device's capabilities in users, which in turn leads to dissatisfaction~\cite{moussawi2020perceptions}.

The split into two high level categories, i) the measurement of factors leading to or arising from anthropomorphism and related concepts, and, ii) explorations of how people conceptualise and understand VAs in relation to humans and social interaction, revealed a contrast in concerns and impacts. For anthropomorphism and personification concerns were clearer and centred around the way these behaviours may change people's interactions with devices. However, the notion that anthropomorphism is a challenge to be solved through design is problematic; given the understanding of these responses as part over-learned social routine, part CASA-style subconscious behaviour, it is not clear that widespread understanding of how voice assistants work would effectively counteract our predisposition to treat them as we treat people.

The deep-seated nature of these responses also challenges the view that anthropomorphism and personification are often associated with incomplete understanding. Where this is based on the language of users to and about VAs, such as the use of gendered pronouns, \citeauthor{festerling2021anthropomorphizing} note the difficulty in determining the \textit{actual} meaning of what people say, recommending that ``future research could be more critical of the role of language in anthropomorphism''~\cite{festerling2021anthropomorphizing}. A related tension surfaces in the literature around CASA and findings that more human-like devices are automatically read by users as more \textit{capable} devices (i.e. they are expected to display human capabilities~\cite{10.1145/3313831.3376789}). However, the potential ethical implications of these questions for work on ontological categories is less clear. Indeterminate results in children and reports of adults switching categories may suggest a new ontological category, but what would constitute an appropriate response from VA designers if this was determined to be true? The literature tended to focus more on the relevance of results to immediate questions of categories without considering the wider implications for HAI, and it seems unlikely that either outcome would lead to changes in the way that VA conversational interfaces are designed; a more likely impact would be changes in how people's responses to VAs are interpreted by researchers.

\subsection{Accessibility}\label{sec:accessibility}
VAs present an interesting combination of opportunities and challenges for accessibility. Their potential user base ranges from those who find it more difficult to navigate visual interfaces, to those who have trouble easily giving verbal commands to VAs and understanding spoken responses. Semi-structured interviews were utilised to discuss frequently used VA functionalities and the issues encountered when accessing them, and this was often supplemented with materials such as recorded videos~\cite{10.1145/3459990.3465195}, and manual interactions with VAs using a pre-collected corpus~\cite{10.1145/3308560.3317597, 10.1145/3234695.3236354, ballati2018hey}. An interesting alternative approach was to indirectly explore how children access and interact with VAs through parent observations, allowing for the inclusion of an additional secondary perspective~\cite{10.1145/2771839.2771910}.

The ability of VAs to accurately transcribe people's voices and communicate smoothly with them was a key theme across the reviewed papers. The accuracy of speech recognition provided by VAs (particularly Google Assistant and Siri~\cite{10.1145/3234695.3236354, ballati2018hey}) was generally viewed favourablely by users, and those with language impairments had an even higher level of approval~\cite{10.1145/3173574.3174033}. However, performance was less satisfactory for children, with VAs appearing to have been designed without a way to properly bridge the gap between children's and adults' expressive language skills. This was frequently frustrating for children~\cite{10.1145/2771839.2771910, 10.1145/3196709.3196772}, and was made worse when they tried unsuccessfully to seek help~\cite{10.1145/3459990.3465195}, or found that the VA could not correctly pronounce their name~\cite{10.1145/3196709.3196772}. Difficulties maintaining conversations were also the subject of investigation, commonly focusing on the short, nonadjustable listening window of VAs which caused problems for children and people with language impairments who may require more time to give a response~\cite{10.1145/2771839.2771910, 10.1145/3196709.3196772, 10.1145/3173574.3174033, 10.1145/3459990.3465195}. Another focus was how people struggled to repair VA conversations, having to rely on external entities (e.g. parents~\cite{10.1145/3196709.3196772}) to maintain dialogues, increasing the difficulty of using VAs~\cite{10.1145/3173574.3174033, 10.1145/3459990.3465195}.

In general there was positive sentiment towards VAs in accessibility contexts. \citeauthor{10.1145/3234695.3236344} argued that VAs were crucial in instilling feelings of independence and empowerment in blind users~\cite{10.1145/3234695.3236344, 10.1145/3439231.3440603}, with similar results related to mobility~\cite{correia2019virtual}. Parents meanwhile, acknowledged that VAs had created many enjoyable moments for their children~\cite{10.1145/3196709.3196772, 10.1145/3459990.3465195}, and if interactions with VAs began at a very young age these experiences could influence the way children interacted with other technologies~\cite{10.1145/3196709.3196772}. However, barriers still remain to true inclusivity, with variable performance across languages, accents, gender, and other demographics~\cite{10.1145/3308560.3317597}. Because these studies focused on highly specific populations, there was little engagement with wider issues of accessibility and few comparisons made across demographics (i.e. it is difficult to gain a clear picture of the complete state of accessibility for any given VA). The continual evolution of the voice models used in VAs introduces an opening for follow-up studies, although we did not encounter any such papers in the survey. Another interesting difference that emerged was in framing: performance differences were treated more as \textit{engineering} problems, in contrast with work on e.g. gender, where similar issues were more often motivated by and framed as social inequity (see Section~\ref{sec:gender}).

\subsection{Social Order}\label{sec:social_order}
A commonly studied dynamic in the literature was between parents and children, where researchers focused on understanding whether VAs could foster parent-child communication and enhance parental practices. On the one hand, using a VA at home gave parents additional opportunities to improve their communication skills, and the features of VAs also helped parents in achieving their parenting goals and promoted parent-child dynamics~\cite{beneteau2020parenting}. On the other, parents complained that they had to strictly regulate both the amount of time children spent using VAs~\cite{beneteau2020parenting} and access to adult content~\cite{storer2020all}, sometimes even having to be physically present with their children during use~\cite{beneteau2020parenting}; deciding when and how to allow children access to voice assistants can therefore be an additional burden for parents~\cite{biele2019might}, which runs counter to the intention of providing a more relaxed parenting environment.

A major concern raised about voice assistants was the extent to which they could entrench or disrupt domestic power structures. Sharing a VA between household members could create tension where cohabitants did not use the device equally~\cite{storer2020all}, with mismatches between willingness to use VAs in shared spaces leading to a reduction in use and eventually abandonment~\cite{trajkova2020alexa}. The use of VAs within established social orders is generally hierarchical and managed in line with existing household social structures, with users negotiating use of the VA when intentions conflicted~\cite{porcheron2018voice, kudina2021alexa}. In this way, voice assistants were seen more as tools integrated into existing power structures than disruptive forces that overturn household social orders. This runs counter to fears (particularly in the news\footnote{\url{https://qz.com/701521/parents-are-worried-the-amazon-echo-is-conditioning-their-kids-to-be-rude}}) that the adoption of voice assistants could destabilise the social order of the home e.g. by answering requests by children that would be considered rude if posed to a person. There \textit{is} a risk, however, that already problematic power imbalances could be exacerbated by VAs, especially when devices are not controlled by the people that use them (e.g. in intimate partner violence and surveillance in offices, student halls, rental accommodation, etc.~\cite{kudina2021alexa, seymour2020strangers, freed2018stalkers, slupska2021threat}).

While analysing the survey papers we found that talking about the social order of the home overlapped with the related concept of group dynamics, whereby the social dynamics of a group are shaped by the emotional state and behaviour of each member~\cite{barsade2007does,10.1145/3171221.3171275}. \citeauthor{lee2020hey} similarly measured how shared device usage affects `group harmony' as a way to measure changes in group dynamics~\cite{lee2020hey}, referring to the ties among group members in terms of mutual support, appreciation, care, emotional attachment, and cooperation~\cite{venter2012impact}. Continued use of VAs appears to have a favourable impact on group harmony through psychological satisfaction and dependence developed by users~\cite{lee2020hey}, especially when this makes participating in family activities more accessible (see Section~\ref{sec:accessibility})~\cite{storer2020all}.

\subsection{Performance of Gender}\label{sec:gender}
A key concern for researchers studying the performance of gender by voice assistants was the feminine presentation of major voice assistants that seems to persist across cultures. This was often justified from a psychological point of view, with studies suggesting that people perceive women's voices and names as sounding more gentle, kind, and caring than men's~\cite{costa2019ai, schneider2021recommended}, with VAs read as female therefore presented as more acceptable to users. In addition, the findings of \citeauthor{10.1145/3411763.3451623}'s empirical research demonstrated that female voices were significantly more trusted in assistance tasks than compliance tasks~\cite{10.1145/3411763.3451623}.

But early findings on CASA suggested that people apply existing gender stereotypes to computers~\cite{nass1994computers}, and these harmful preconceptions (e.g. around women's roles in society and the types of jobs they should perform) have influenced the creation of current norms and expectations for the performance of gender by VAs. Given the prevalence of white men on engineering and design teams, the role of VAs in reflecting and reinforcing these stereotypes is the source of intense discussion~\cite{costa2019ai}. In the past, women were expected to perform a variety of stereotypically gendered labour, such as placing orders, giving reminders, seeking information, taking notes and making calls~\cite{schiller2019alexa, sutko2020theorizing}. Feminising VAs was seen as a reflection of male designers' psychological needs and tendencies---engineers tend to create artefacts that fit within their own social spaces---strengthening the connection between women and submissiveness and satisfying other `heterosexual fantasies'~\cite{walker2020alexa}. 

Given that gendered presentation can be problematic, a series of studies with controlled experiments~\cite{10.1145/3411763.3451623} and interviews~\cite{10.1145/3405755.3406123} explored the factors that cause people to gender devices based on voice and questions around gender-ambiguous voices. Conclusions were polarised. Some findings suggested that investigating voices exhibiting gender ambiguity was worthwhile as gender-ambiguous voices are perceived similarly to gendered voices, and thus do not impact user's trust in VAs~\cite{10.1145/3411763.3451623}. On the other hand, synthesised voices designed to be genderless (e.g. Q\footnote{\url{https://www.genderlessvoice.com}}) are often coded as male or female by listeners, with \citeauthor{10.1145/3405755.3406123} finding that people have specific gender expectations that make this kind of gendering automatic when hearing VA voices~\cite{10.1145/3405755.3406123}. Including genderless voices can itself be problematic if they take the approach of smoothing out differences in voices rather than acknowledging and representing diversity~\cite{10.1145/3449206}. Q in particular has been criticised for drawing distinctions between trans and male/female voices, as well as presenting trans and non-binary voices as a monolithic mid-point of the binary it is attempting to break free from~\cite{10.1145/3449206}. As a way forward, it has been suggested that voice assistants could be designed to randomly choose a voice or switch between them~\cite{habler2019effects}, but this is not the only cue that influences perceptions of gender in VAs. Other design elements such as the physical appearance of devices/interfaces, product branding, specific pronunciations in the speech, and the activity that the VA is currently performing are also influential~\cite{10.1145/3405755.3406123}. This ties in with gendered preferences for voice assistants, particularly around trust, privacy, ease of use, and mobile self-efficacy~\cite{nguyen2019integrated}.

\subsection{Accountability, Conflicts of Interest, Misinformation, and Transparency}
As the above concerns were represented by only six papers between them, we briefly summarise them together here. The one paper coded as accountability measured the efficacy of the certification process for Alexa and Google Assistant skills/actions, finding that 100\% and 39\% of policy-violating skills were certified by the respective platforms~\cite{10.1145/3372297.3423339}. Another paper discussed the inherent conflicts of interest built into VAs, whereby assistants appeared to be acting in users' best interests whilst also prioritising information and services that benefit vendors (e.g. through shopping platforms)~\cite{aguirre2020ai}. Of the three papers on misinformation, two focused on the accuracy of information available through popular devices~\cite{alagha2019evaluating, dambanemuya2021auditing} and one on the inefficacy of spoken warnings alongside content identified as misinformation~\cite{sharevski2021two}. While requests for information about vaccines were handled reasonably well by Google Assistant and Siri, Alexa understood fewer queries and was less likely to present information from authoritative sources~\cite{sharevski2021two}. For news queries, Alexa returned more relevant and timely information, but subtle changes in question phrasing led to significant changes in the relevance and source of information~\cite{dambanemuya2021auditing}. The paper on transparency closely linked this concern to privacy, claiming that modern encryption mechanisms hamper transparency around data collection by requiring secret symmetric keys (i.e. between assistants and vendors, which users cannot access)~\cite{flikkema2017things}.

\section{Discussion \& Future Work}\label{sec:discussion}
\subsection{Who are Voice Assistants Designed For?}
The survey highlights several areas where the interests and needs of the people using VAs fall secondary to those of their manufacturers: data collection for tracking and advertising, the prevalence of female-coded voices as the default, rigid interaction and access control models that are not aligned with inter- and intra-household use, and the lack of unprofitable adjustments to allow more universal access. Some of these design decisions, such as the preference for female-coded voices, originate from the first commercially available voice assistants; as Siri and Alexa did, others followed. In other cases like the neglect of multi-user use, VAs were more likely shaped by data protection regulations that are modelled around the relationship between individual data subjects and corporate data controllers. Finally, issues like poor voice recognition performance for non-native speakers are likely the result of expectations set by the limitations of early voice recognition technologies, designers creating products that work optimally for people like themselves, and the perceived expense of achieving more equitable recognition.

As the technology and expertise required to develop voice assistants and skills become more accessible, it is important that these legacy design decisions are not unthinkingly perpetuated by the devices of the future. Evaluating voice assistants against previous guidelines for human-AI interaction~\cite{10.1145/3290605.3300233} shows that some findings of the literature review are specific instances of wider problems with AI systems. \citeauthor{10.1145/3290605.3300233} find that contemporary VAs are close to meeting some of these guidelines, such as G5 (match relevant social norms) and G7 (support efficient invocation), but the results of the literature survey show that they fall short of others like G6 (mitigate social biases) and G11 (make clear why the system did what it did). Extending this work to produce guidelines that are specific to VAs represents an excellent opportunity to move beyond current design norms and `reset' assumptions around how voice assistants should operate.

\subsection{Widening Participation in VA Research}
A clear theme when conducting the literature review was the community's focus on voice assistants used in Western countries; for example, despite the existence of many Chinese and Korean language VAs, none of the studies reviewed recruited participants resident in China and only one recruited participants from South Korea. This is surprising given China's population and the existence of well-known voice assistant brands in the country. Some of these assistants support multiple dialects, suggesting the potential for shared insights across these VAs and e.g., work on accessibility. There is also a risk that participant recruitment is seen as an opportunity to reduce potential variables, at the expense of making the field representative. Revisiting the observation from Section~\ref{sec: privacy} that Western legal and cultural norms have strongly shaped the evolution of associated voice assistants, we are not aware of an analysis of the influences on VAs outside of the US and Europe---this constitutes an important piece of future work.

As a result of the above---and as evidenced by Section~\ref{sec:whoconcerns}---there is a clear lack of diversity amongst those who participate in research on voice assistants. We therefore take this opportunity to present a challenge and provocation for voice assistant researchers: given the recent upheavals to the way that we work and do research, there is no excuse for a field that proudly pursues diversity to continue to exclude those who live outside a handful of wealthy Western countries. The increasing reach of crowdsourcing platforms commonly used in the survey papers such as Mechanical Turk and Prolific Academic significantly lowers the barrier for data collection with under-represented demographics~\cite{van2022intersectional}, and the quality of the research produced during the coronavirus pandemic demonstrates that collecting qualitative data over the internet is more viable than previously thought. Outside of Western platforms, many others exist that offer diverse participant bases and localisation services to facilitate participation across language and cultural borders. This could be done by not restricting participants based on geography/nationality or, where language is important to the research questions being investigated, including comparative analysis between e.g. native/non-native speakers. This will both broaden the applicability of results as well as identify exclusionary factors that would otherwise go undiscovered.

\subsection{Deepening Explorations of Gender}
Throughout the survey, issues around gender repeatedly surfaced around how gender is performed by VAs, and how they often seem designed for men (e.g. by having lower accuracy for other voices). Norms around gender can be so tightly woven into home and social structures that the introduction of a device that performs gender \textit{and} affects work done in the home inevitably causes disruption. While it is promising to see initial work on voices around and beyond the gender binary, less focus is given to the gendered effects of voice recognition accuracy and how their design affects existing household relationships and power structures shaped by gender.

Beyond the diversification of design teams and training corpora, there are several approaches that can be taken. Providing more nuanced and inclusive representations of gender is a matter of corporate social responsibility rather than solely a design decision. This is most easily achieved by providing more than one voice, and by not labelling voices by gender (e.g., Google Assistant labels voices with colours). The reviewed literature also highlights the large difference in the role that gender plays in interpersonal and human-computer interactions now compared to when the foundational work in this space was undertaken almost 30 years ago. As such, it is important that we revisit these early studies and their implications from a contemporary perspective.

\subsection{The Effects of Habituation}
When looking at longer-term usage trends of VAs there is a discussion over the ways that usage changes over time; usage appears to stabilise after an initial playful phase~\cite{10.1145/3196709.3196772}, but there is a lack of data available on usage trends beyond the scale of days~\cite{10.1145/3311956} or months~\cite{10.1145/3196709.3196772}. One question that arises when trying to contextualise results on privacy perceptions and VAs is the extent to which user perceptions and behaviours will change over a longer period of time. Work relating to categories hints at shifting perspectives around humanness and machine-ness~\cite{10.1145/2470654.2466455}, which may cause related changes towards other aspects of VAs. Larger-scale changes in cultural and commercial attitudes to data collection by devices and `creepy' functionality are also likely to manifest in user perceptions, and it may be that as the gap widens between contemporary and early research on voice assistants that researchers need to take care when comparing their results with prior work. An opportunity therefore exists to re-run existing studies to determine how perceptions might be changing in different cultures.

A related longer term aspect of novelty concerns the transparency and accountability of voice assistants as they evolve and become integrated into more devices in the home and beyond. Speculative work on the future of voice assistants~\cite{10.1145/3357236.3395479} imagines futures with ubiquitous voice assistants where people give commands to be answered by whichever assistant is present, \textit{without necessarily knowing who created or controls that assistant}. A key fear raised in this speculative work pertains to undisclosed functionality, where users are unpleasantly surprised by the VA's inferential abilities and the real world effects that the VA can cause. As it becomes the norm to have voice control built into consumer electronics and commoditisation increases the feasibility of assistants from smaller vendors, it will become increasingly important to know which assistant is being used at any given time and (more importantly) the associated capabilities, limitations, and interests involved. Mandatory use is also raised as a concern, which echoes the discussion in the accessibility literature (Section~\ref{sec:accessibility}).

\subsection{A Shift in the Human-VA Relationship?}
The main conclusion from the analysis of the reviewed papers and subsequent discussion seems to be that VAs are quickly becoming a ubiquitous presence in people's lives. While initially a curiosity, most people with smartphones now have access to a voice assistant. One way that this ubiquity manifests is in the extension of existing platforms and services, with VAs changing the way these are accessed to make them seamlessly available throughout the home (e.g., music and search). This transition comes naturally as people are already familiar with e.g. Spotify, and so the choice to use it via a VA quickly becomes subconscious.

When considering the long term impact of voice assistants one can draw parallels with how the smartphone drastically changed the ways that people relate to one another and the patterns for social interaction. While smartphones are inherently mobile and thus extend interactions outside of the home in ways not previously possible, voice assistants primarily change the way that people interact with digital technology \textit{within} the home. By becoming a persistent part of the home environment, voice assistants subtly change the way that we interact with each other in the home. The rigid interaction and family models built into these devices constrains the social interactions that people share with others, and can cause social friction. Clear examples of this arise around cohabitation and managing users within and between family units where VAs do not adhere to existing norms around those relationships and concerns over parenting~\cite{goulden2021delete}---in this sense, VAs constrain people's ability to be a partner/roommate/parent in artificial ways. 

Another ready comparison with smartphones is the ability to opt-out of owning and using the technologies whilst continuing to participate in society. The disruption caused by smartphones has ushered in ``a new way of living wherein the smartphone is ordinary, necessary, and integral''~\cite{10.1145/2470654.2466134}, one where the key decision is whether to \textit{own} the device. With VAs the opposite is often true---the packaging of voice assistant software with new smartphones, TVs, and headphones means that a huge number of people already have access to a voice assistant, making the choice one of use rather than ownership. This could make it easier for VAs to become a necessary or default means of interacting with digital platforms and services in the future.

\subsection{Unexpected Discoveries}\label{sec:discoveries}
While preliminary work on the background literature and prior work had suggested that anthropomorphism was a major category of ethical concern, analysis of the included papers revealed a web of related but distinct concepts that extend beyond this relatively narrow classification. As a result, this concern was renamed `social interaction' to better reflect the range of research questions that deal with how people interact with voice assistants (differentiated from social order concerns that focus on how voice assistants affect relationships with other people). This also opens up the range of potential research questions to include a wider variety of social interactions and, as VAs become more sophisticated, the different ways in which we might communicate and build trust with VAs beyond simple task-oriented interactions.

Another unanticipated class of concern emerged around misinformation. A common discussion point around other devices and platforms that facilitate access to information, voice assistants present an unfortunate collection of attributes that make them particularly apt to perpetuate misinformation. Not only does voice as a medium heavily promote short, easy to understand interactions, it also makes it difficult to provide information on sources and links to further reading. There is potential for companion smartphone apps and displays built into smart speakers to introduce more nuance to fact-finding, but their efficacy depends on users interacting with a secondary modality after initiating a verbal search for information with the assistant. Other avenues of exploration could include the verification of fact sources and mandatory communication to users about the source of a skill's information before and during use. Exploration of this topic represents an exciting opportunity for future work, but will be made difficult by the dominant architecture where skills are hosted by third party developers outside of the control of vendors (resulting in difficulties when vetting and verifying third party software) with very recent work showing evidence of third-party skills serving misinformation~\cite{bispham2023}.

\subsection{Limitations}
We struck a balance with the databases we searched between accurately representing the literature, volume of results, and ease of running complex searches. Running searches in English across English-speaking venues inevitably influenced the literature returned, but at the same time major publishers describe themselves as global institutions and we note that many region specific conferences such as ACM's Asia CCS use English as their working language. The addition of hand-picked papers that evaded the systematic searches will also have influenced the results, but these represented less than 2\% of the total number of papers screened and were carefully balanced to maximise the coverage of the review. Despite this, there will also have been in-scope papers that were not included in the analysis.

When classifying metadata we did not distinguish between a work's target and effective demographic (e.g., studies that did not set out to examine particular groups but recruited from pools with known demographic biases like university students). The same applies to the small number of cases where the country of residence was not reported and was thus coded as the platform used for recruitment (the U.S. represents \textasciitilde4\% of the world's population but almost half of Mechanical Turk workers~\cite{paolacci2010running}).

\section{Conclusion}
We systematically reviewed 117 research papers on ethical concerns with VAs, consolidating the incredible work done by the community. We highlight areas of consensus, disagreement, and gaps in the body of knowledge that can guide future research, and consider the distribution of approaches and methods across the field. Our findings show that some concerns like privacy were much more likely to be operationalised for quantitative research than others like accessibility, and that the people participating in these studies are overwhelmingly from North America and Europe. We outline key areas to be addressed by future work, such as widening participation and revisiting early results from a contemporary perspective, with the hope of making future VAs more equitable and inclusive.

\begin{acks}
This research was funded by the UK Engineering and Physical Sciences Research Council under grant EP/T026723/1.
\end{acks}

\bibliographystyle{ACM-Reference-Format}
\bibliography{main}

\section*{Appendix: Statistical Results and List of Reviewed Papers}

\begin{table*}[h]
    \centering
    \begin{tabular}{r|l}
        \toprule
        Sub-population & \# of Papers \\ \hline
        None & 55 \\
        Children & 10 \\
        Blind and visually impaired & 3 \\
        People with Dysarthria & 2 \\
        Older adults & 2 \\
        Trans and non-binary & 1 \\
        Users with disabilities & 1 \\
        Users with motor impairments & 1 \\
        Portuguese Speakers & 1 \\
        Young Adults (18-36) & 1 \\
        \bottomrule
    \end{tabular}
    \caption{Number of papers recruiting specific groups, written as reported.}
    \label{tab:subpops}
\end{table*}

\begin{table*}[h]
    \centering
    \begin{tabularx}{0.76\paperwidth}{r|X}
    \toprule
        Primary Concern & Works Included \\ \hline
        Accessibility & \cite{10.1145/2771839.2771910, 10.1145/3234695.3236354, 10.1145/3196709.3196772, 10.1145/3234695.3236344, ballati2018hey, 10.1145/3173574.3174033, 10.1145/3308560.3317597, correia2019virtual, 10.1145/3405755.3406129, 10.1145/3368426,10.1145/3439231.3440603, brause2020externalized, 10.1145/3459990.3465195} \\
        Accountability & \cite{10.1145/3372297.3423339} \\
        Social Interaction & \cite{10.1145/3027063.3053246, 10.1145/3176349.3176868, 10.1145/3209626.3209709, 10.1145/3290605.3300488, 10.1145/3359316, 10.1145/3308532.3329466, 10.1145/3290605.3300772, 10.1145/3338286.3340116, yuan2019speech, wagner2019human, 10.1145/3313831.3376665, 10.1145/3383652.3423906, 10.1145/3404983.3405513, moussawi2020effect, benlian2020mitigating, festerling2020alexa, aeschlimann2020communicative, ghosh2020understanding, ernst2020impact, 10.1145/3459990.3460730, moussawi2020perceptions, whang2021like, moriuchi2021empirical, 10.1145/3479515, carolus2021alexa} \\
        & \cite{lee2021social, 10.1145/3290607.3310422, 10.1145/3469595.3469607, belk2018morphing} \\
        Conflict of Interest & \cite{aguirre2020ai} \\
        Gender & \cite{schneider2021recommended, loideain2020alexa, schiller2019alexa, sutko2020theorizing, nguyen2019integrated, 10.1145/3340764.3344441, 10.1145/3405755.3406123, 10.1145/3411763.3451623, 10.1145/3449206, costa2019ai, walker2020alexa, adams2019addressing} \\
        Misinformation & \cite{alagha2019evaluating, dambanemuya2021auditing, sharevski2021two} \\
        Privacy & \cite{10.1145/3170427.3188448, 10.1145/3274371, 10.1145/3290605.3300669, 10.1145/3311956, 10.1145/3313831.3376529, 10.1145/3313831.3376551, 10.1145/3320269.3384732, 10.1145/3339252.3340330, 10.1145/3369807, 10.1145/3375188, 10.1145/3411170.3411260, 10.1145/3411764.3445122, 10.1145/3412383, 10.1145/3427228.3427250, 10.1145/3427228.3427277, 10.1145/3428361.3428469, 10.1145/3449119, liao2019understanding, chalhoub2020alexa, konrad2020right, pal2020personal, vimalkumar2021okay, lutz2021privacy, easwara2015privacy, phipps2021your, lin2020transferability, furey2019can, kalin2019amaze, burbach2019hey, germanos2020privacy, huxohl2019interaction, pridmore2020personal, malkin2019privacy, abdi2019more, seymour2020strangers, dubois2020speakers, cobb2021would, bolton2021security, sharif2020smart, sahu2019challenges, zeng2019understanding, abrokwa2021comparing, krueger2020using, mccarthy2020shouting, ha2021exploring, park2021users, voorveld2020social, anniappa2021security, kennedy2019can, naraparaju2020fingerprinting, krueger2020using} \\
        Social Order & \cite{porcheron2018voice, lee2020hey, trajkova2020alexa, biele2019might, beneteau2020parenting, storer2020all, kudina2021alexa} \\
        Transparency & \cite{flikkema2017things} \\
        \bottomrule
    \end{tabularx}
    \caption{Complete list of papers included in the review.}
    \label{tab:all-concerns}
\end{table*}

\begin{table*}[h]
    \centering
    \begin{tabular}{c|r|r|r|r|r}
    \toprule
    Concern & Primary (All) & Primary (Quant) & Primary (Qual) & Primary (Theory) & Primary (Mixed) \\
    \hline
    Privacy & 50 & 29 & 14 & 2 & 5 \\
    Social Interaction & 29 & 12 & 12 & 2 & 3 \\
    Accessibility & 13 & 2 & 9 & 0 & 2 \\
    Gender & 12 & 4 & 1 & 7 & 0 \\
    Social Order & 7 & 1 & 5 & 1 & 0 \\
    Misinformation & 3 & 2 & 0 & 0 & 1 \\
    Accountability & 1 & 1 & 0 & 0 & 0 \\
    Conflicts of Interest & 1 & 0 & 0 & 1 & 0 \\
    Transparency & 1 & 0 & 0 & 1 & 0 \\
    \hline
    Concern & Secondary (All) & Secondary (Quant) & Secondary (Qual) & Secondary (Theory) & Secondary (Mixed) \\
    \hline
    Privacy & 9 & 2 & 5 & 2 & 0 \\
    Social Interaction & 7 & 4 & 3 & 0 & 0 \\
    Social Order & 5 & 0 & 4 & 1 & 0 \\
    Gender & 3 & 1 & 2 & 0 & 0 \\
    Autonomy & 3 & 0 & 2 & 0 & 1 \\
    Transparency & 2 & 0 & 1 & 1 & 0 \\
    Accessibility & 2 & 0 & 2 & 0 & 0 \\
    Accountability & 1 & 0 & 1 & 0 & 0 \\
    None & 89 & 44 & 25 & 10 & 10 \\
    \bottomrule
    \end{tabular}
    \caption{Primary and secondary concern by approach. Note that papers may have zero or mulitiple secondary concerns.}
    \label{tab:concerns-stats}
\end{table*}
\end{document}